\documentclass[twocolumn,dvipdfmx]{aastex6}

\usepackage{color} 

\shorttitle{TMC-1}
\shortauthors{Dobashi et al.}

\begin{document}


\title{Spectral Tomography for the Line-of-Sight Structures of the Taurus Molecular Cloud 1}


\author{
Kazuhito Dobashi\altaffilmark{1},Tomomi Shimoikura\altaffilmark{1}, \\
Fumitaka Nakamura\altaffilmark{2,3}, Seiji Kameno \altaffilmark{2,4},
Izumi Mizuno \altaffilmark{5,6,7}, and Kotomi Taniguchi \altaffilmark{3,5}
}

\affil{\scriptsize{\rm $^1$ dobashi@u-gakugei.ac.jp}}
\altaffiltext{1}{Department of Astronomy and Earth Sciences, Tokyo Gakugei University, Koganei, Tokyo  184-8501, Japan}
\altaffiltext{2}{National Astronomical Observatory of Japan, Mitaka, Tokyo 181-8588, Japan}
\altaffiltext{3}{The Graduate University for Advanced Studies (SOKENDAI), 2-21-1 Osawa, Mitaka, Tokyo 181-0015, Japan}
\altaffiltext{4}{Joint ALMA Observatory, Alonso de C\'{o}rdova 3107 Vitacura, Santiago, Chilie}
\altaffiltext{5}{Nobeyama Radio Observatory, National Astronomical Observatory of Japan 462-2 Nobeyama, Minamisaku, Nagano 384-1305, Japan}
\altaffiltext{6}{Department of Physics, Faculty of Science, Kagoshima University, 1-21-35 Korimoto, Kagoshima, Kagoshima 890-0065, Japan}
\altaffiltext{7}{East Asian Observatory, 660 N. A'oh\={o}k\={u} Place, University Park, Hilo, HI 96720, USA}
 

\begin{abstract}
We clarify the line-of-sight structure of the Taurus Molecular Cloud 1 (TMC-1)
on the basis of the CCS($J_N=4_3-3_2$) and HC$_3$N($J=5-4$) spectral data observed at
a very high velocity resolution and sensitivity of $\Delta V \simeq 0.0004$ km s$^{-1}$ ($=61$ Hz) and
$\Delta T_{\rm mb} \simeq 40$ mK. The data were obtained toward the cyanopolyyne peak
with $\sim$30 hours integration using the Z45 receiver and the PolariS spectrometer installed
in the Nobeyama 45m telescope.
Analyses of the optically thin $F=4-4$ and $5-5$ hyperfine lines of the HC$_3$N emission
show that the spectra consist of four distinct velocity components
with a small line width ($\lesssim 0.1$ km s$^{-1}$)
at $V_{\rm LSR}=$5.727, 5.901, 6.064, and 6.160 km s$^{-1}$, which we call
A, B, C, and D, respectively, in the order of increasing LSR velocities.
Utilizing the velocity information of the four velocity components,
we further analyzed the optically thicker CCS spectrum and the other hyperfine lines of the
HC$_3$N emission by solving the radiative transfer to investigate how the four velocity components
overlap along the line of sight. Results indicate that they are located in the order of A, B, C, and D
from far side to near side to the observer, indicating that TMC-1 is shrinking, moving inward
as a whole. \\
\end{abstract}


\section{INTRODUCTION} \label{sec:intro}

Molecular emission lines provide us with essential information on the kinematics of molecular clouds.
Especially, emission lines from low rotational transitions of carbon monoxide
\citep[such as the $J=1-0$ transition of CO, e.g., ][]{Dame2001}
have been widely used in various studies of molecular clouds and star formation.
Combination of the optically thick $^{12}$CO line and the optically much thinner lines from its
isotopologues ($^{13}$CO and C$^{18}$O) are powerful to
reveal to the spatial distributions of molecular gas in a wide range of density
from diffuse outskirts to dense cores forming young stars.

The C$^{18}$O emission line is generally optically thin, and is useful to investigate dense
regions.
It is even possible to quantify the velocity field of dynamically infalling gas
toward young stars \citep[e.g.,][]{Ohashi1997,Momose1998,Shimoikura2016}.
However, we often face difficulties to reveal structures
of such dense regions only with the C$^{18}$O emission line,
because it has a rather low critical density ($\sim10^{2-3}$ cm$^{-3}$),
and thus its spectra tend to be contaminated by a large amount of emission
from the diffuse background and foreground along the line-of-sight (LOS).
In addition, the C$^{18}$O molecules tend to be depleted onto dust grains in cold and dense regions
\citep[e.g.,][]{Bergin2002},
which makes a dip in the C$^{18}$O spectra unrelated to the velocity field.
The density and velocity distributions around the densest regions in molecular clouds
are of our particular interest in terms of star formation, which should be 
better probed by other molecular lines with much higher critical densities.
If we use two or more such molecular lines with different optical depths,
it should also be possible to reveal the structure of dense regions along the LOS.

In this paper, we introduce a useful method to probe into the LOS structures of
dense regions in molecular clouds, and report results of its applications to
the Taurus Molecular Cloud 1 (TMC-1).
The method utilizes the combination of the CCS$(J_N=4_3-3_2)$
and HC$_3$N$(J=5-4)$ emission lines at 45 GHz whose
critical densities are $\sim 10^{4-5}$ cm$^{-3}$.
The basic idea of the method is to identify distinct velocity components
based on the optically thin hyperfine lines of the HC$_3$N emission,
and to determine the order of the detected components along the LOS
by solving the radiative transfer to reproduce the observed CCS emission line
as well as the other hyperfine lines of the HC$_3$N emission
which are much optically thicker and thus sensitive to the
absorption by the gas in the foreground.

TMC-1 is a well-known filamentary condensation and a part of
an extended molecular cloud in Taurus. 
In this paper,  we will refer to the condensation
as the TMC-1 filament, or more simply, TMC-1.
Carbon-chain molecules such as CCS and HC$_3$N
are known to be abundant in TMC-1 and in other
young star forming regions
\citep[e.g.,][]{Hirahara1992,Suzuki1992,Hirota2009,Shimoikura2012,
Taniguchi2016a,Taniguchi2016b,Taniguchi2017a,Taniguchi2017b,TaniguchiSaito2017},
and thus the emission lines from these molecules can be a useful tool to investigate
the initial conditions of star formation.
In addition, the CCS and HC$_3$N emission lines at 45 GHz are relatively intense
compared to those from other carbon-chain molecules, and they also have an
advantage to be observable simultaneously with a millimeter-wave radio telescope
like the 45m telescope at the Nobeyama Radio Observatory (NRO),
which we actually used to obtain the data for this paper,
because their rest frequencies differ only by $\sim10$ MHz.

The CCS and HC$_3$N spectral data used in this paper
were obtained using the NRO 45m telescope.
The data were obtained for the purpose to detect the Zeeman splitting
of the CCS line to measure the magnetic field in TMC-1.
About 30 hours integration provided us with wonderful spectral data
having an extremely high velocity resolution of 0.0004 km s$^{-1}$.
Detection of the Zeeman splitting and the related analyses of the magnetic
field will be presented in a subsequent publication (Nakamura et al. 2018, in preparation).
In this paper, we will concentrate on the analyses of the LOS structures of TMC-1
inferred from the spectral data.

We describe the observational procedures in Section \ref{sec:obs}.
In Section \ref{sec:results}, we will explain how we analyzed the spectral data,
and present results of the analyses. We have detected four distinct velocity components,
and found that they are lying on the LOS from far side to near side to the observer
in the order of increasing radial velocities, suggesting the inward motion of TMC-1.
In Section \ref{sec:dis}, we further discuss the global inward motion of TMC-1 on the
basis of the $^{13}$CO and C$^{18}$O data available on the website of NRO.
Summary of this paper is given in Section \ref{sec:conclusions}.


\section{OBSERVATIONS} \label{sec:obs}

Observations were made with the 45m telescope at NRO for 8 days
in the period between 2015 April to 2016 April.
We used the dual-polarization receiver Z45 \citep{Nakamura2015}
which enabled us to observe the CCS($J_N=4_3-3_2$) and HC$_3$N($J=5-4$) emission
lines simultaneously. The receiver provided a typical system noise temperature
of $T_{\rm sys}=100-200$ K. The beam size of the telescope
was $\sim37\arcsec$ (HPBW) at 45 GHz, and the pointing accuracy was better than $\sim5\arcsec$
during the observations, as we checked by observing the SiO maser NML Tau
every $\sim1.5$ hours.

\begin{figure*}
\begin{center}
\includegraphics[scale=0.8]{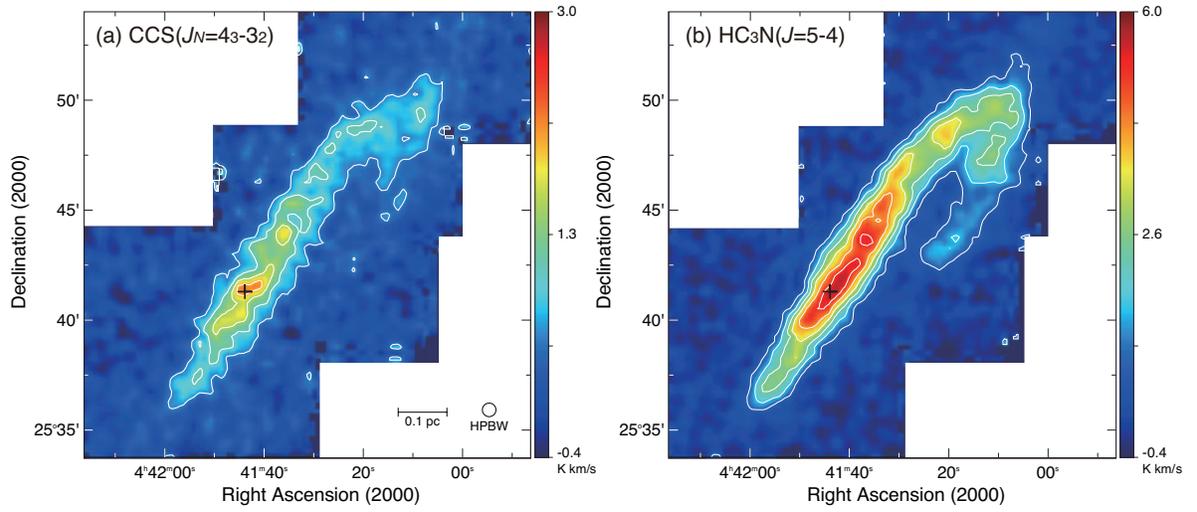}
\caption{Integrated intensity maps of the (a) CCS($J_N=4_3 - 3_2$) and (b) HC$_3$N($J=5-4$) emission
lines. In the HC$_3$N map, only the main line (consisting of $F=6-5$, $5-4$, and $4-3$) is integrated.
The data are converted to $T_{\rm mb}$ by dividing the original $T_{\rm a}^*$ data
by the main beam efficiency of the 45m telescope ($\eta_{\rm mb}=0.7$ at 45 GHz).
The lowest contours and contour intervals are
$\int T_{\rm mb} dv=$ 0.5 K km s$^{-1}$ for the CCS map and 0.8 K km s$^{-1}$for the HC$_3$N map.
Plus sign denotes the cyanopolyyne peak.
\label{fig:intensity_maps}}
\end{center}
\end{figure*}

In the beginning of the observations, we made an On-The-Fly (OTF) mapping
covering the entire extent of the TMC-1 filament with the SAM45 spectrometer \citep{Kamazaki2012}
which covers a velocity range of $\sim200$ km s$^{-1}$ with a velocity resolution
of $\sim 0.05$ km s$^{-1}$ ($=7.63$ kHz).
The purpose of the mapping is to determine the CCS intensity peak position more precisely than
the old measurements done with a simple position-switching mode \citep{Hirahara1992}.
Resulting maps are shown in Figure \ref{fig:intensity_maps}.
As indicated by the plus sign in the maps,
we measured the coordinates of the CCS peak position to be
$\alpha_{\rm J2000}$=04$^{\rm h}$41$^{\rm m}$43.87$^{\rm s}$ and 
$\delta_{\rm J2000}$=+25$^{\circ}$41$\arcmin$17.7$\arcsec$,
which is well consistent with the intensity peak position
of the CCS$(J_N=4_3-3_2)$ emission line
revealed by \citet[][see their Figure 9]{Wolkovitch1997},
and we performed the long-time integration toward this position.
As the backend for the long-time integration,
we used the special digital spectrometer PolariS \citep{Mizuno2014} which covers
two $\sim4$ MHz bandwidths with an extremely
high frequency resolution of 61 Hz corresponding to a velocity resolution of 0.0004 km s$^{-1}$ at 45 GHz.
The integration was performed by the position switching mode using the emission-free OFF position
offset from the target ON ponsition by $(\Delta \alpha,\Delta \delta)=(30\arcmin,0\arcmin)$.
For the integration, we employed the Smoothed-Bandpass Calibration (SBC) technique \citep{Yamaki2012}
which greatly shortens the total observing time. 

The standard chopper-wheel method \citep{Kutner1981} was employed to scale the spectral data to units of $T_{\rm a}^*$,
and the daily intensity fluctuation of the spectra was better than $10$ $\%$ during the observations.
Because the main beam efficiency $\eta_{\rm mb}$ at the observed frequency (45 GHz) was not
established precisely in the beginning of the observations, we processed the spectral data in this paper
without converting to the $T_{\rm mb}$ scale. The value of $\eta_{\rm mb}$ was better established later to be
$\eta_{\rm mb}=70 \pm 2$ $\%$ at 45 GHz, which can be used to scale the spectra presented in this paper.
In total, we performed the integration of $\sim30$ hours, and suppressed the noise level to $\Delta T_{\rm a}^*\simeq40$ mK
at the 0.0004 km s$^{-1}$ velocity resolution.

We summarize parameters of the observed molecular lines in Table \ref{tab:45obs},
and show sample spectra taken with the SAM45 spectrometer in Figure \ref{fig:sam45}.
As shown in the table and figure, the HC$_3$N$(J=5-4)$ emission consists of five hyperfine lines,
and three of them ($F=6-5$, $5-4$, and $4-3$) are optically thick and blended, which we call ``main line'' in this paper.
The other two hyperfine lines ($F=4-4$ and $5-5$) are optically thin and isolated from the other lines, which we call
``satellites''. The CCS line is moderately but not very optically thick and does not have hyperfine lines.



\section{RESULTS \& ANALYSES}\label{sec:results}

\subsection{Basic Assumptions}
We summarize basic assumptions and equations that we use
throughout this section to analyze the observed CCS and HC$_3$N emission lines.

We first assume that the observed spectra consist of $N(=1,2,3, ...)$ velocity components,
and that they are lying at different positions along the LOS as indicated
by $i$ $(=0, 1, 2,..., N-1$) in Figure \ref{fig:assumptions}. 
We assign $i=0$ to the furthest position from the observer.

We also assume that the antenna temperature of the $i$-th component $T_{\rm a}^i$,
which would be observed when there is only the $i$-th component along the LOS,
is expressed as a function of the radial velocity $V$ by the following well-known formula,
\begin{equation}
\label{eq:radiative_trans}
T_{\rm a}^i(V) = T_0^i [ 1-e^{-\tau^{i}(V)}]
\end{equation}
where $\tau^{i}(V)$ is the optical depth of the $i$-th component that we will explain later in this subsection.
$T_0^i $ is related to the excitation temperature $T_{\rm ex}^i$ as
\begin{equation}
\label{eq:radiative_trans2}
T_0^i = \eta_{\rm mb} \left[{J_{\rm RJ}}(T_{\rm ex}^i) - {J_{\rm RJ}}({T_{\rm bg}})\right] \, 
\end{equation}
where $T_{\rm bg}$ and $\eta_{\rm mb}$ are
the cosmic background ($T_{\rm bg}=2.725$ K)
and the main beam efficiency of the 45m telescope ($\eta_{\rm mb}=0.70 \pm 0.02$ at 45 GHz).
We assume $T_{\rm ex}^i$ to be
constant at all velocities for the $i$-th component,
but it can change depending on the molecular lines, because the CCS and HC$_3$N lines
should be emitted from similar but slightly different density regions in TMC-1.
$J_{\rm RJ}(T)$ is an effective temperature equivalent to that in the Rayleigh-Jeans law,
and is expressed as
\begin{equation}
\label{eq:Jrj}
{J_{\rm RJ}}(T) = \frac{{h\nu /k}}{{{e^{h\nu /kT}} - 1}} \,
\end{equation}
where $h$, $k$, and $\nu$ are the Planck constant, Boltzmann constant,
and the rest frequency of the molecular lines, respectively.

In addition, we assume that the distributions of the radial velocities of the molecular gas
in each velocity component follows a single Gaussian function, and thus the optical depth
of the $i$-th component $\tau^i (V)$ in Equation (\ref{eq:radiative_trans})
also follows a single Gaussian function, because the optical depths are
proportional to the total number of molecules along the LOS for a constant excitation temperature.

\begin{figure}
\begin{center}
\includegraphics[scale=0.8]{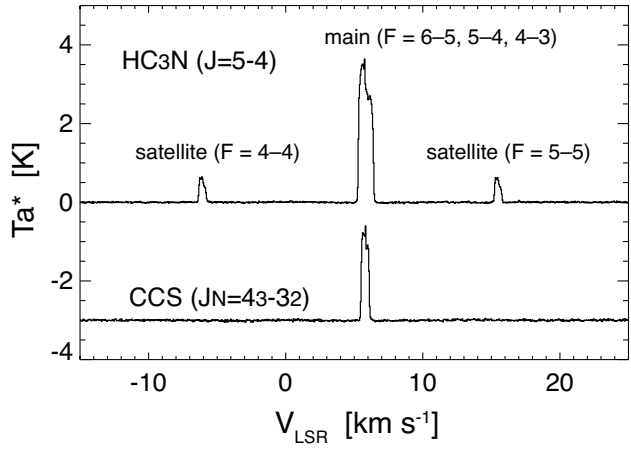}
\caption{The HC$_3$N($J=5-4$) and CCS($J_N=4_3-3_2$) spectra taken toward TMC-1 with the SAM45 spectrometer.
The CCS spectrum is offset by $-3$ K.
\label{fig:sam45}}
\end{center}
\end{figure}

\begin{figure*}
\begin{center}
\includegraphics[scale=0.6]{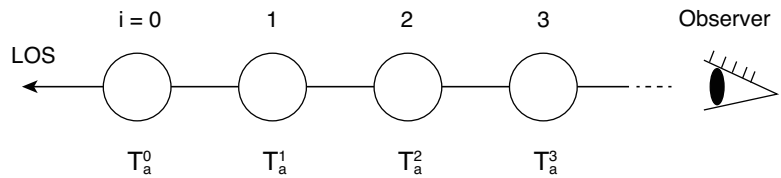}
\caption{
Schematic illustration for the locations of the velocity components
along the LOS, indicated by different values of $i$ $(=0, 1, 2, 3...)$.
We assign $i=0$ to the furthest location from the observer.
$T_{\rm a}^i$ is the antenna temperature of the $i$-th velocity component expressed by Equation (\ref{eq:radiative_trans}),
which should be observed when there is only the $i$-th component along the LOS.
Spectrum that should be actually observed is the sum of $T_{\rm a}^i$ scaled by a factor of $e^{-S^i}$
as expressed in Equation (\ref{eq:Ta}).
\label{fig:assumptions}}
\end{center}
\end{figure*}

In the case of the CCS line,
$\tau^i (V)$  can be simply expressed as
\begin{equation}
\label{eq:tau_ccs}
\tau ^i(V) = \tau_0^i \, {{\rm exp}\left[{ - \frac{1}{2}{{\left( {\frac{{V - V_0^i}}{{\sigma ^i}}} \right)}^2}}\right]}\
\end{equation}
where $\tau_0^i$ , $V_0^i$, and $\sigma^i$ are the maximum optical depth, the peak radial velocity,
and the velocity dispersion, respectively.
Note that the above equation can be applied also to each of the satellites of the HC$_3$N line, because, 
as seen in Figure \ref{fig:sam45}, they are well separated from the other hyperfine lines.

In the case of the main line of HC$_3$N,
$\tau ^i(V)$ needs to be calculated as the sum of the optical depths of the individual hyperfine lines.
For the spectrum shown in Figure \ref{fig:sam45} including the satellites,
$\tau ^i(V)$  can be expressed as
\begin{equation}
\label{eq:tau_hc3n}
\tau^i(V) = {\sum\limits_{j = 0}^4 {R_j} \, \tau _{\rm 0}^i \, {\rm exp} {\left[ - \frac{1}{2}\left( {\frac{{V - V_0^i - V_f^j}}{{\sigma^i }}} \right)^2 \right] }}
\end{equation}
where $R_j$ is the normalized line intensity of the $j$-th
hyperfine line  (i.e., $\sum R_{j}=1$), and
$\tau _{\rm 0}^i$ is the total optical depth of the HC$_3$N($J=5-4$) line
for the $i$-th velocity component.
$V_f^j$ is the ostensible velocity shift arising from the choice of the
rest frequency when converting the frequency axis to the radial velocity axis using the Doppler effect.
In Table \ref{tab:45obs}, we summarize the values of  $R_{j}$ and $V_f^j$
for the case when we set the rest frequency to that of the $F=5-4$ hyperfine line.

Finally, we assume that the beam filling factors of all the $N$ velocity components are 1. 
In other words, we assume that the velocity components are shading one another completely
within the beam of the 45m telescope. In that case, the total optical depth between
the $i$-th component and the observer ${S^i}(V)$
can be written as 
\begin{equation}
\label{eq:total_tau}
{S^i}(V) = \left\{ \begin{array}{l}
\sum\limits_{k= i + 1}^{N - 1} {\tau^k(V) \,\,\, ({\rm for} \, i \ne N - 1)} \\
\,\,\,\,\,\,\,\,\,\,\,\,\,\,\, 0 \,\,\,\,\,\,\,\,\,\,\,\,\,\,\,\, ({\rm for} \, i = N - 1) \,.
\end{array} \right.
\end{equation}
Using ${S^i}(V)$ in the above, the antenna temperature $T_{\rm a}^*(V)$
that we should observe toward TMC-1 can be modeled as
the sum of $T_{\rm a}^i(V)$
in Equation (\ref{eq:radiative_trans})
scaled by a factor of $e^{ - {S^i}(V)}$ as
\begin{equation}
\label{eq:Ta}
T_{\rm a}^*(V) = \sum\limits_{i = 0}^{N - 1} {{T_{\rm a}^i}(V)} {e^{ - {S^i}(V)}} \,\, .
\end{equation}

According to the measurements by \citet{Suzuki1992} and \citet{Hirahara1992},
the maximum optical depths of the CCS line and the HC$_3$N line
in TMC-1, i.e., $\tau_0^i$ in Equations (\ref{eq:tau_ccs}) and (\ref{eq:tau_hc3n}),
are an order of $\sim 1$ and $\sim 10$, respectively, inferring that the optical depth
of the satellites of the HC$_3$N emission (calculated as $R_j \tau_0^i$ with
$R_j=0.0134$ in Table \ref{tab:45obs}) is an order of $\sim0.1$.
In the following subsections, we will first analyze the optically thin satellites of the HC$_3$N emission
to identify distinct velocity components.
Using the information of the
identified components, we will further analyze the optically thicker CCS line and
the main line of the HC$_3$N emission by fitting the observed spectra
with Equation (\ref{eq:Ta})
to investigate how the identified velocity components are lying along the LOS.



\subsection{Analyses of the Satellite Lines of HC$_3$N} \label{sec:satellites}

\begin{figure}
\begin{center}
\includegraphics[scale=0.8]{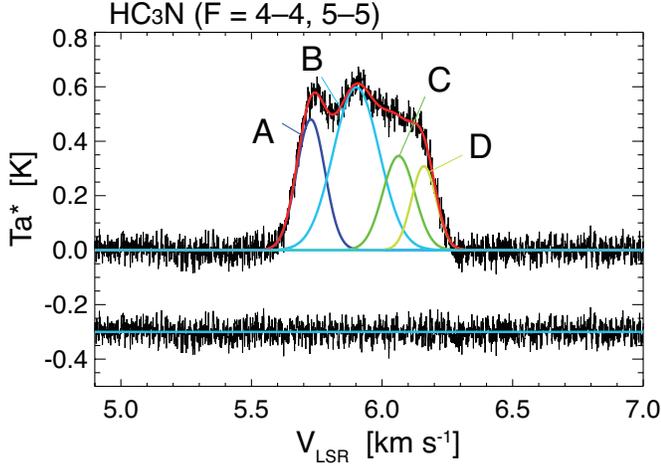}
\caption{The averaged spectrum of the $F=4-4$ and $5-5$ lines of the HC$_3$N($J=5-4$) emission
obtained with the PolariS spectrometer. The noise level and the velocity resolution of the spectrum are
$\Delta T_a^*=$0.0385 K and 0.0004 km s$^{-1}$. The spectrum is smoothed to the 0.0008 m s$^{-1}$
velocity resolution in the figure. The red line denotes the model best fitting the observed spectrum.
The resulting reduced $\chi^2$ is 1.0016.
Spectra shown by different colors with the labels from A to D are the four velocity components
expressed in Equation (\ref{eq:satellite}). 
The residual offset by $-0.3$ K is shown below the spectrum.
\label{fig:F4455}}
\end{center}
\end{figure}

As expected, the $F=4-4$ and $5-5$ lines of the HC$_3$N emission agree perfectly
with each other in shape and intensity,
because they have the same line intensities (i.e., $R_j$ for $j=0$ and $4$ in Table \ref{tab:45obs}). 
We thus averaged them to reduce the noise.
Figure \ref{fig:F4455} shows the averaged spectrum.
We assume that the line can be expressed as a simple sum of $N$-component
Gaussian functions,
because the line should be optically thin ($\tau \sim 0.1$ at most)
as mentioned in the above.
For such an optically thin case, Equation (\ref{eq:Ta}) with $\tau^i$
in Equation (\ref{eq:tau_ccs}) can be
approximated to
\begin{equation}
\label{eq:satellite}
T_{\rm a}^*(V) = \sum\limits_{i = 0}^{N - 1} {T_{\rm s}^i} \, {{\rm exp}  \left[ { - \frac{1}{2}{{\left( {\frac{{V - V_0^i}}{{\sigma^i}}} \right)}^2}} \right]}
\end{equation}
where $T_{\rm s}^i$ is the peak antenna temperature of the $i$-th component.

In order to find the most appropriate value of $N$,
we fitted the averaged spectrum with Equation (\ref{eq:satellite}) for increasing $N$ from 1,
and investigated how the equation can fit the spectrum well for different values of $N$.
We found that the reduced $\chi^2$ of the fit (hereafter, $\chi_{\rm r}^2$) decreases rapidly up to $N=4$,
and remains around $\chi_{\rm r}^2 =$1 for $N \ge 4$.
For example, we obtained 
$\chi_{\rm r}^2$ = 2.6543, 1.0896, 1.0016, and 1.0013
for $N=$2, 3, 4, and 5, and we found that the $i$-th component for $i \ge 5$ tend to fit the apparent
spurious noise in the spectrum.
We therefore conclude that the satellite lines of HC$_3$N
consist of $N=4$ velocity components.
These velocity components have a narrow line width of
$\sigma^i \simeq 0.05-0.09$ km s$^{-1}$
and are separated by $\sim 0.1$ km s$^{-1}$ in the radial velocity.
Hereafter, we will call the four velocity components A, B, C, and D in the order of
increasing radial velocities.
We show the best model in Figure \ref{fig:F4455},
and summarize the model parameters in Table \ref{tab:satellite}.

We should note that \citet{Wolkovitch1997} observed
TMC-1 in CCS and showed results of
Gaussian fit for a spectrum taken at a
position $\sim1\arcmin$ away from the position we observed (see their Figure 10).
Though they assumed $N=3$ velocity components, their values for the
radial velocities and velocity dispersions are similar to what
we derived here, and we regard that the two datasets are basically consistent within
the uncertainties of the adopted main beam efficiencies and the assumed rest frequencies
of the observed lines (see Section \ref{sec:ccs}).


\subsection{Analyses of the CCS Line \label{sec:ccs}}

\begin{figure}
\begin{center}
\includegraphics[scale=0.8]{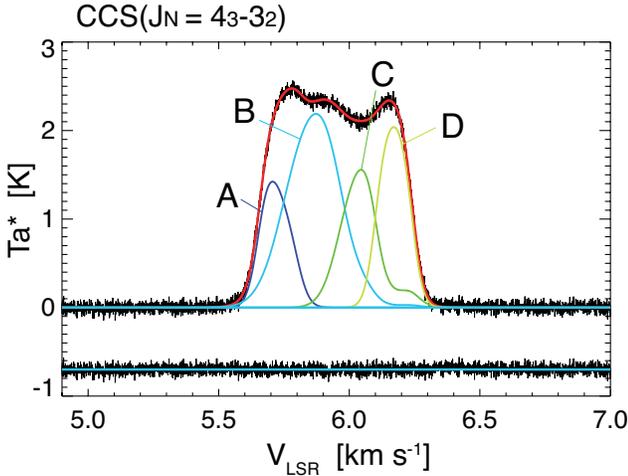}
\caption{The CCS($J_N=4_3-3_2$) spectrum
obtained with the PolariS spectrometer.
The noise level and the velocity resolution of the spectrum are
$\Delta T_a^*=$0.0417 K and 0.0004 km s$^{-1}$.
The red line denotes the model best fitting the observed spectrum.
The resulting reduced $\chi^2$ is 1.0631.
The four velocity components labelled from A to D 
are the term $ {{T_{\rm a}^i}(V)} {e^{ - {S^i}(V)}}$ in
Equation (\ref{eq:Ta}).
The residual offset by $-0.7$ K is shown below the spectrum.
\label{fig:ccs}}
\end{center}
\end{figure}

Figure \ref{fig:ccs} displays the CCS line obtained with the PolariS spectrometer.
Because the critical densities of the
HC$_3$N and CCS lines are more or less similar ($n$[H$_2$]$\sim10^{4-5}$),
they should trace a similar density region in TMC-1, and
we would expect that the CCS line should consist of the same velocity components
as those identified in the satellite lines of HC$_3$N.

However, the CCS line is not optically thin, and we cannot reproduce the CCS spectrum
in Figure \ref{fig:ccs} as a simple sum of the velocity components as expressed
in Equation (\ref{eq:satellite}). 
We need to decide the positions for the four velocity components A--D
along the LOS (i.e., $i$ in Figure \ref{fig:assumptions}), and
to fit the observed CCS spectrum with Equation (\ref{eq:Ta}).

There are $N!=4!=24$ permutations to locate the four components A--D
to the positions $i=0-3$.
For each permutation, we fitted the observed CCS spectrum in Figure \ref{fig:ccs}
by Equation (\ref{eq:Ta}), leaving $T_0^i$ in Equation (\ref{eq:radiative_trans}) as well as
$\tau_0^i$ and $\sigma^i$ in Equation (\ref{eq:tau_ccs}) as free parameters,
and calculated the resulting $\chi^2$. 

In the fitting, we basically fixed the peak velocities $V_0^i$ to the
values measured from the satellite lines of HC$_3$N (Table \ref{tab:satellite}).
We should note, however,
that there is an uncertainty of 20 kHz
in the rest frequency of the CCS line (Table \ref{tab:45obs}),
corresponding to the $\sim0.1$ km s$^{-1}$ uncertainty in velocity.
Comparing the model and observed spectrum,
the uncertainty apparently causes an artificial shift in velocity.
For this problem, we set one more free parameter $V_{\rm shift}$
and replaced $V-V_0^i$ on the right side of Equation (\ref{eq:tau_ccs}) by $V-V_0^i-V_{\rm shift}$
to perform the fitting.

The best model with the minimum $\chi_{\rm r}^2$($=1.063$) is found when the components (A, B, C, D)
are located at the positions $i=$(0, 1, 2, 3) in this order. We show the best model in Figure \ref{fig:ccs}
and summarize the model parameters in Table \ref{tab:ccs}.
The results that the components with lower
radial velocities are located further from the observer is striking, because it infers
the global inward motion of TMC-1 (see Figure \ref{fig:model}). However, the inferrered order of the four velocity
components along the LOS (i.e., A, B, C, D for $i=0$, 1, 2, 3) is not a unique solution at this stage,
because there are many other permutations of the four velocity components that can give
a good fit to the CCS spectrum with $\chi_{\rm r}^2$ values lower than the 90 $\%$ confidence level ($\chi_{\rm r}^2 =1.152$).
For example, the orders of (A, B, D, C), (B, A, C, D), (D, A, C, B), and (D, C, A, B) for $i=$(0, 1, 2, 3) gives secondary
best fits with $\chi_{\rm r}^2 \simeq1.081$.


\subsection{Analyses of the Main Line of HC$_3$N \label{sec:main}}

\begin{figure}
\begin{center}
\includegraphics[scale=0.8]{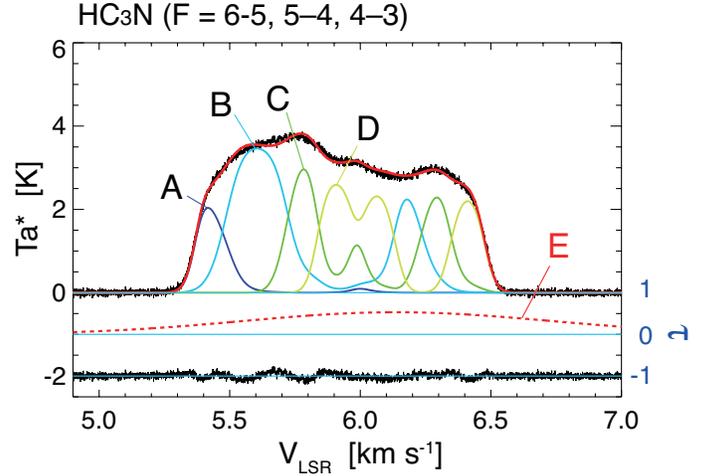}
\caption{The main component of the HC$_3$N($J=5-4$) emission line
obtained with the PolariS spectrometer. The noise level and the velocity resolution of the spectrum are
$\Delta T_a^*=$0.0459 K and 0.0004 km s$^{-1}$.
The red line denotes the model best fitting the observed spectrum.
The resulting reduced $\chi^2$ is 2.2093.
The four velocity components labelled from A to D 
are the term $ {{T_{\rm a}^i}(V)} {e^{ - {S^i}(V)}}$ in
Equation (\ref{eq:Ta}).
The residual offset by $-2$ K is shown below the spectrum.
The  optical depth of the additional component E is shown by the red broken line.
\label{fig:main}}
\end{center}
\end{figure}

To better establish the order of the components A--D along the LOS,
we attempted to fit the main line of HC$_3$N shown in Figure \ref{fig:main}
in the same way as for the CCS spectrum. However, we found that
the observed spectrum cannot be fit well only with the four components,
and the resulting best $\chi_{\rm r}^2$ remains in the range $20-30$ for
any permutations of A--D.
This is mainly because, while we could nicely fit half of the spectrum in the lower velocity range
($V_{\rm LSR} \lesssim 5.8$ km s$^{-1}$), the other half in the higher velocity range
($5.8 \lesssim V_{\rm LSR} \lesssim 6.6$ km s$^{-1}$) is apparently weaker
by $\sim10$ $\%$ than what we would expect from its intensity in the
lower velocity range.
Because the SBC technique is employed, this mismatch should not be
due to contamination by the emission from the OFF position.

After some trials, we finally concluded that we need one more velocity
component which is special in a sense that it doesn't
radiate the emission but contribute only to the absorption
in the higher velocity range.
We believe that the additional component should be the diffuse molecular
gas in the outskirts of TMC-1, which is lying between the four components A--D
and the observer and is not dense enough to excite the HC$_3$N line.
The absorption by the diffuse gas in the foreground should be serious in the case of
the HC$_3$N main line, because the line is optically thick
as we will show in the following. The same absorption must be occurring
also in the CCS emission, but the effect should be small because the line is
optically much thinner ($\tau \lesssim 0.1$ for the CCS line).
 
\begin{figure*}
\begin{center}
\includegraphics[scale=0.6]{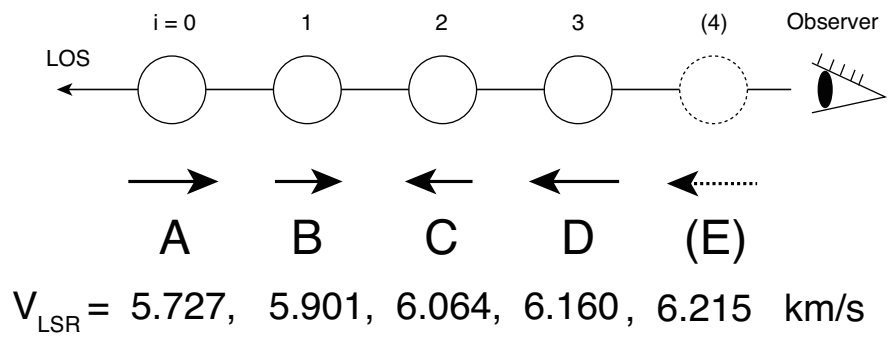}
\caption{
Schematic illustration for the locations of the velocity components
A, B, C, D, and E along the line of sight. Arrows indicate their relative motions
inferred by our analyses.
\label{fig:model}}
\end{center}
\end{figure*}

As illustrated in Figure \ref{fig:model},
we call the additional component ``E'', and assign the fixed position $i=4$
(i.e., the closest to the observer) along the LOS to model the observed
HC$_3$N main line.

In the same way as for the CCS line, we assume that the original emission
from the $i$-th component of the HC$_3$N line can be expressed
in the form of Equation (\ref{eq:radiative_trans}).
Here, we set $T_{\rm 0}^4=0$ K for the component E,
because it doesn't radiate the emission as explained in the above.

The optical depth of the $i$-th component including E $\tau ^i(V)$
should follow Equation (\ref{eq:tau_hc3n}), and
we take that $V_0^i$ and $\sigma^i$ for A--D
are the same as those found for the satellite lines (Table \ref{tab:satellite}).

For $N=5$ components including the component E,
the total optical depth between the $i-$th component and the observer ${S^i}(V)$ and 
the observed antenna temperature $T_{\rm a}^*(V)$ of the HC$_3$N main line
should be modeled by Equations (\ref{eq:total_tau}) and (\ref{eq:Ta}), respectively.
The way to fit the observed main line of the HC$_3$N emission
is basically the same as for the CCS line, but note that  
the position of the component E along the LOS is fixed to $i=4$,
and it doesn't radiate the emission ($T_{0}^4=0$ K)
and contribute only to the absorption ($\tau_{0}^4 \ne 0$).

Under these restrictions, we fitted the observed HC$_3$N spectrum in Figure \ref{fig:main}
with Equation (\ref{eq:Ta}) in the same way as for CCS. 
In the fitting, we left seven parameters as free parameters, i.e.,
$V_0^4$, $\sigma^4$, $\tau_0^4$ (parameters for E),
and $T_0^i$ for $i=0-3$ (parameters for A--D).
We fixed $V_0^i$ and $\sigma^i$ for the components A--D to the values obtained by
fitting the satellite line (Table \ref{tab:satellite}). Also, we didn't leave $\tau_0^i$ for A--D as free parameters,
but calculated them for given $T_0^i$ using the values of $T_{\rm s}^i$ in Table \ref{tab:satellite} as
\begin{equation}
\label{eq:estimate_tau_m0}
\tau _0^i = \frac{{T_{\rm s}^i}}{{{R_0}T_0^i}}
\end{equation}
where $R_0=0.0134$ (i.e., the value of $R_j$ for the $F=4-4$ or $5-5$ transitions in Table \ref{tab:45obs}). 

We repeated the fitting for each of the $4!=24$ permutations to locate the components A--D to the positions $i=0-3$
(Figure \ref{fig:model}), and calculated the $\chi_{\rm r}^2$ to search for the best model.
As a result, we found that the permutation (A, B, C, D) for $i=$(0, 1, 2, 3) gives the best fit to the observed spectrum
with $\chi_{\rm r}^2=2.209$. We should note that the best $\chi_{\rm r}^2$ values for 
all of the other permutations are much higher ($\sim3.1-15.2$)
and are far above the 90 $\%$ confidence level ($\sim2.283$),
which ensures that the order of the components A--D along the LOS
can be uniquely determined.

We display the best model by the red line in Figure \ref{fig:main},
and summarize the obtained model parameters in Table \ref{tab:main}.
As seen in the figure, the fit is not as good as that for the CCS line.
This is apparently because the HC$_3$N main line is optically much thicker than the CCS line
and the shape of the spectrum can be affected more easily by a small variation of
the excitation temperature and density. 

We note that the component E is optically much thinner and has much broader
line width compared to the other components A--D (Table \ref{tab:main}),
which is plausible because the component E should be tracing more turbulent and diffuser gas
surrounding the dense regions of TMC-1.
Such diffuse envelope around the dense
regions can be traced in other molecular lines with a lower critical density (e.g., $^{13}$CO),
which is useful to check and validate our analyses.  
We will discuss this point in Section \ref{sec:dis}.


\subsection{Estimate of the Excitation Temperature \label{sec:Tex}}
Finally, we derived the excitation temperatures $T_{\rm ex}$ of the CCS and HC$_3$N lines in a standard way
\citep[e.g.,][]{Shimoikura2012}.
$T_{\rm ex}$ is related to $T_0^i$ as in Equations (\ref{eq:radiative_trans2}) and (\ref{eq:Jrj}).
We calculated $T_{\rm ex}$ of the CCS and HC$_3$N lines
by inserting the measured values of $T_0^i$ to the equations.
Results are summarized in the last columns of Tables \ref{tab:ccs} and \ref{tab:main}.

As can be seen in the tables,
there is an uncomfortable mismatch in $T_{\rm ex}$ of a few Kelvin
between the two molecular lines, and $T_{\rm ex}$ for HC$_3$N is
always higher than those for CCS for all of the velocity components A--D.
Though the values of $T_{\rm ex}$ for the CCS and HC$_3$N lines do not need to match precisely
because they should be tracing slightly different density regions,
we believe that the difference is caused mainly by the ambiguity in our assumption on
the additional component E.
In our present model, we assume that the diffuse component E has a simple
Gaussian velocity distribution and lies in the foreground of the other components A--D.
However, such diffuse components can exist at every interface of the other components (e.g., between B and C
in Figure \ref{fig:model}), and in addition, their velocity distribution may not follow a simple Gaussian
function. All of these possible effects are not taken into account in our simple model, but
they can cause rather large errors in the analyses of the optically thick HC$_3$N main line,
especially when we determine $T_0^i$ for the line.

On the contrary, the CCS line may be less affected by the diffuse component(s) in the foreground,
because the optical depth of such diffuse gas should be much smaller for the CCS line than for the HC$_3$N main line.
The average ratio of the optical depth of the CCS line to that of the HC$_3$N line in the components A--D
is $\sim0.2$. If we assume the same ratio in the component E, the optical depth of the CCS line would be $\sim 0.1$
for the component E, which means that the derived $T_{\rm ex}$ for CCS
could be underestimated by $\sim 1$ K.


\section{Discussion}\label{sec:dis}

\subsection{Global Inward Motion of TMC-1}\label{sec:infall}
As illustrated in Figure \ref{fig:model}, the analyses in Section \ref{sec:results} strongly indicate
the inward motion of the four velocity components A--D. The additional
diffuse component E in the foreground follows the same trend as the other components,
suggesting the possibility that the entire system of TMC-1 is moving inward, possibly shrinking.
It is noteworthy that the component E has a larger velocity dispersion
($\sim 0.5$ km s$^{-1}$, Table \ref{tab:main})
compared to the other components whose dispersions  ($\sim 0.05$ km s$^{-1}$) are much smaller
and close to the thermal dispersion ($\sim 0.025$ km s$^{-1}$).
Because the component E and the other components should trace the outer and inner regions of TMC-1, respectively,
the difference in the velocity dispersion may infer that
the diffuse gas in the outer region contracts as it loses turbulence
to form more quiescent and denser inner region.

\begin{figure}
\begin{center}
\includegraphics[scale=0.8]{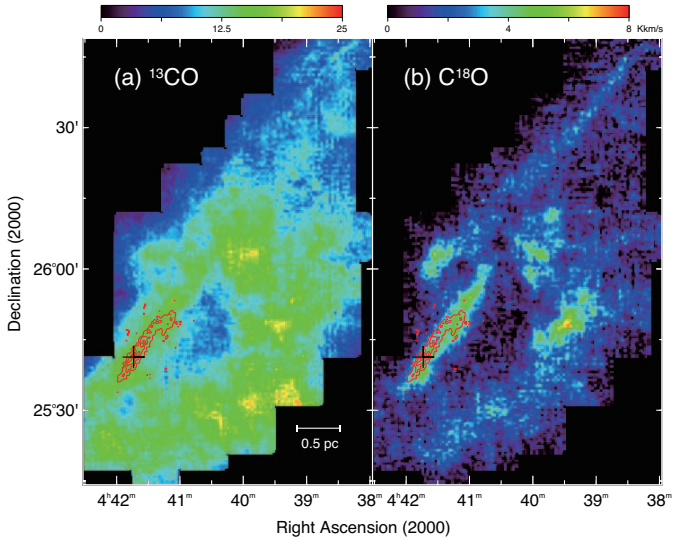}
\caption{
Integrated intensity maps of the (a) $^{13}$CO($J=1-0$) and (b) C$^{18}$O($J=1-0$) emission lines.
The data are converted to $T_{\rm mb}$ by dividing the original $T_{\rm a}^*$ data
taken from the data archive of NRO
by the main beam efficiency of the 45m telescope ($\eta_{\rm mb}=0.45$ at 110 GHz).
Red contours represent the CCS intensity drawn at the same levels as in Figure \ref{fig:intensity_maps}(a).
Plus sign denotes the cyanopolyyne peak.
\label{fig:13co_c18o_maps}}
\end{center}
\end{figure}

The global structure of TMC-1 including the diffuse outer region would be better traced
by emission lines of more abundant molecules with lower critical densities such as
the isotopologues of carbon monoxide.
Figures \ref{fig:13co_c18o_maps} displays the total intensity maps
of  the $^{13}$CO($J=1-0$) and C$^{18}$O($J=1-0$) emission lines,
which we downloaded from the data archive of NRO
\footnote{http://www.nro.nao.ac.jp/~nro45mrt/html/results/data.html}.
As seen in the figure, the emission lines are detected over much larger
regions than CCS and HC$_3$N , tracing lower density regions around TMC-1.
We smoothed the $^{13}$CO and C$^{18}$O data to the same angular resolution as our
CCS and HC$_3$N data, and extracted $^{13}$CO and C$^{18}$O 
spectra at the cyanopolyyne peak, which we display in Figure \ref{fig:13CO_C18O}(a).
These spectra provide us with various information on the internal structures of TMC-1.
For example, the $^{13}$CO emission is evident over the velocity range
where the component E is inferred to exit ($\sim6-7$ km s$^{-1}$),
ensuring that there is rich molecular gas over the velocity range.
In addition, the $^{13}$CO spectrum is characterized by the asymmetric profile
with slightly higher temperature at lower velocity (indicated by the arrow labelled ``X" in the figure) than
at higher velocity  (``Y"). We point out that such feature can be seen only when
there are gradients both in the velocity and in the temperature along the LOS,
i.e., the feature can be observed only when TMC-1 is moving inward and having a 
higher temperature in the inner region, or when TMC-1 is moving outward and having a 
lower temperature in the inner region. The line profile of the former case is similar
to that of infalling cores forming young stars at the center \citep[e.g.,][]{Zhou1992,Zhou1993}.
In the case of TMC-1, there is no known heating source (i.e., young stars) at the center,
and it is rather puzzling that the temperature is higher in the inner region than in the outer region,
because we would expect the opposite temperature distribution for such a dense cloud in general.
As the difference in temperature is small being only a few Kelvin, it could be merely due to
random fluctuation of the temperature in TMC-1.

We further point out that there is a bump in the $^{13}$CO spectrum
at $V_{\rm LSR}\simeq 7$ km s$^{-1}$ (indicated by the arrow labelled ``Z"),
and the bump is faintly detected also in C$^{18}$O.
We are not sure about the origin of the bump, but it may be the emission from a condensation
in the foreground or background, not directly related to the TMC-1 filament.

Finally, we point out that there is a clear dip in the C$^{18}$O spectrum shown in Figure \ref{fig:13CO_C18O}(a) over the
velocity range where we detected the CCS emission.
This may be due to the depletion of C$^{18}$O onto dust grain, which is often
observed in dense cloud interiors \citep[e.g.,][]{Bergin2002}.

\subsection{A Model for the Inward Motion}
\begin{figure}
\begin{center}
\includegraphics[scale=0.8]{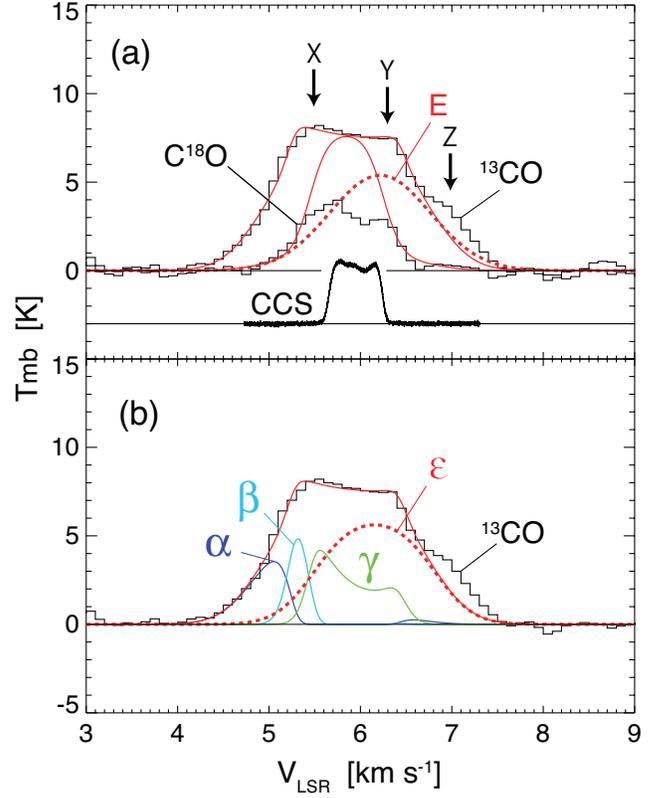}
\caption{Black solid lines represent the $^{13}$CO and C$^{18}$O
spectra taken from the data archive of NRO, which are compared with the
spectra calculated using the model in Figure \ref{fig:illust}.
(a) Thin solid red lines represent the $^{13}$CO and C$^{18}$O spectra
calculated with model parameters best fitting the observed spectra
(see Section \ref{sec:13comodel}).
The spectrum labelled ``E''  shown by the red broken line is the additional component
(see Section \ref{sec:main} and Figure \ref{fig:main}).
The C$^{18}$O, $^{13}$CO, and CCS spectra are scaled to units of $T_{\rm mb}$
assuming the main beam efficiency of $\eta_{\rm mb}= 0.45$, $0.45$, and $0.70$, respectively.
The CCS spectrum is offset by $-3$ K.
The intensity scale of the component E is arbitrary.
(b) Thin red solid line represents the calculated $^{13}$CO spectrum.
Contributions of the four parts labelled $\alpha$, $\beta$, $\gamma$, and $\varepsilon$
in Figure \ref{fig:illust} are shown by
the lines with different colors. Note that the 
emission from $\varepsilon$ (red broken line) is very similar to the additional component E.
\label{fig:13CO_C18O}}
\end{center}
\end{figure}

{\label{sec:13comodel}}
Taking into account the above inspection, we attempt to model TMC-1 in the following
to reproduce the observed $^{13}$CO spectrum in Figure \ref{fig:13CO_C18O}(a),
which should help our better understanding of the global inward motion of TMC-1
including the diffuse outer region traced in $^{13}$CO.
We assume that TMC-1 is a cylindrical filament
consisting of dense quiescent inner region
surrounded by
diffuse turbulent outer region,
and regard that the identified velocity components A--D are representing dense substructures
of the inner region, which we call ``subfilaments" in this paper.

\begin{figure}
\begin{center}
\includegraphics[scale=0.70]{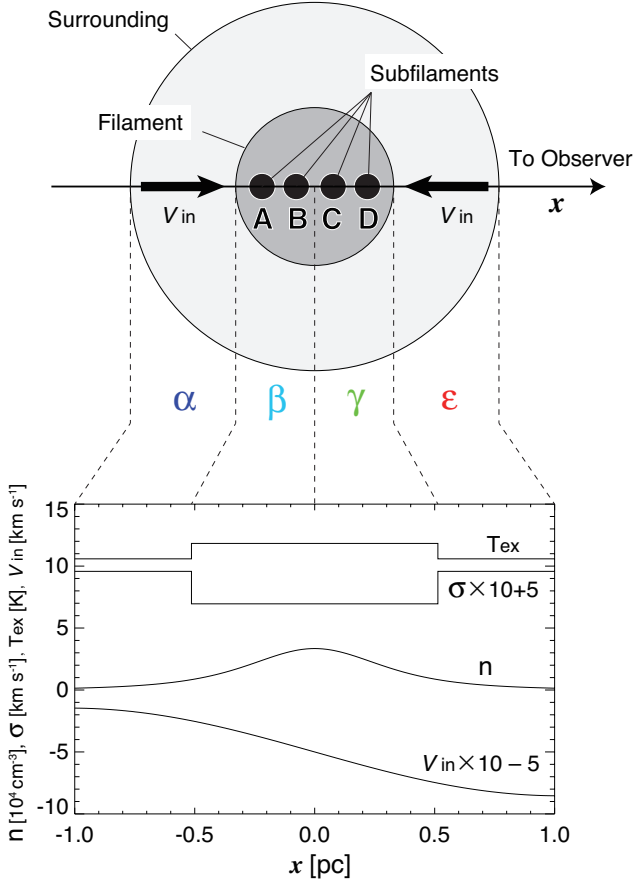}
\caption{Schematic illustration of the internal structure of TMC-1.
We assume a cylindrical filament whose cross section is shown in the
upper panel. The filament consists of the diffuse outer region with larger turbulence
and the dense inner region with less turbulence. The dense subfilaments A--D are
embedded in the inner region.
We set the $x$ axis passing through the center of the filament
and pointing toward the observer.
Four solid lines in the lower panel denote the distributions of the parameters
of the filament
$n$, $\sigma$, $T_{\rm ex}$, and $V_{\rm in}$
in Equations (\ref{eq:density}), (\ref{eq:sigma}), (\ref{eq:T}), and (\ref{eq:Vin}), respectively,
best fitting the observed $^{13}$CO and C$^{18}$O spectra
as shown in Figure \ref{fig:13CO_C18O}.
In the panel, $\sigma$ and $V_{\rm in}$ are scaled by 10,
and offset by $+5$ km s$^{-1}$ and $-5$ km s$^{-1}$, respectively.
We divide the outer and inner regions into
$\alpha$, $\beta$, $\gamma$, and $\varepsilon$ at $|x|=R_0$
as denoted in the figure in order to investigate their contributions to the resulting
spectra (see Figure \ref{fig:13CO_C18O}).
\label{fig:illust}}
\end{center}
\end{figure}

Figure \ref{fig:illust} illustrates the intersection of the model filament.
As shown in the figure, we set the $x$ axis passing through the center of the filament and pointing
toward the observer, along which
we calculate the
$^{13}$CO spectrum expected to be observed $T_{\rm e}$
as a function of the radial velocity $V$ in the same way as we did in Sections \ref{sec:ccs} and \ref{sec:main},
but using the following equations;

\begin{equation}
\label{eq:Te}
{T_{\rm e}}(V) = \int_{ - \infty }^{ + \infty } {T(x)} \left[ {1 - {e^{ - \tau (x,V)}}} \right]{e^{ - S(x,V)}}dx \,\, ,
\end{equation}
\begin{equation}
\label{eq:Tx}
T(x) = \eta_{\rm mb} \left[{J_{\rm RJ}}(T_{\rm ex}(x)) - {J_{\rm RJ}}({T_{\rm bg}})\right] \, ,
\end{equation}
\begin{equation}
\label{eq:tau}
\tau (x,V) = {\tau _0}(x) \, {{\rm exp} \left[ { - \frac{1}{2}{{\left( {\frac{V-V_{\rm in}(x)}{\sigma(x)}} \right) }^2}} \right]} \, ,
\end{equation}
and
\begin{equation}
\label{eq:S}
S(x,V) = \int_x^{ + \infty } {\tau (y,V)dy}  \,\, 
\end{equation}
where $V_{\rm in}(x)$ is the inward velocity,
and
$\sigma(x)$ and $T_{\rm ex}(x)$ are the velocity dispersion
and excitation temperature of $^{13}$CO.
$J_{\rm RJ}$ is a function given in Equation (\ref{eq:Jrj}), and
$\tau (x,V)$ and $S (x,V)$ are the optical depth per unit length and
the total optical depth to the observer. 
${\tau _0}(x)$ is the peak optical depth
that can be derived from
the number density
of molecular hydrogen
$n(x)$,
the excitation temperature $T_{\rm ex}(x)$,
and the fractional abundance of $^{13}$CO \citep[e.g.,][]{Shimoikura2013}.
In this paper, we assume the fractional abundance
to be
$[^{13}{\rm CO}]/[{\rm H_2}]=2\times10^{-6}$ \citep{Dickman1978}.

Here, we set the model parameters $\sigma(x)$,
$T_{\rm ex}(x)$,
${V_{\rm in}}(x)$,
and $n(x)$ which is equivalent to ${\tau_0}(x)$.
For simplicity, we assume that $n(x)$ follows the density law
for an isothermal cylindrical cloud with infinite length and in the gravitational equilibrium \citep{Stodolkiewicz1963,Ostriker1964,Inutsuka1992},
which can be expressed as

\begin{equation}
\label{eq:density}
n(x) = {n_0}{\left[ {1 + {{\left( {\frac{x}{{{R_0}}}} \right)}^2}} \right]^{ - 2}}
\end{equation}
where $n_0$ is the density at $x=0$ .
$R_0$ is the effective radius which
is related with $n_0$ as 
\begin{equation}
\label{eq:R0}
R_0 = \sqrt {\frac{2}{{\pi G \overline{\mu}  {m_{\rm H}}{n_0}}}} {c_{\rm s}}
\end{equation}
where $G$, $\overline{\mu}$, $m_{\rm H}$, and $c_{\rm s}$ are the gravitational
constant, mean molecular weight ($\mu_{\rm H}=2.8$), proton mass,
and speed of sound in the filament.

We divide the filament into the inner and outer regions at $R_0$, and assume
that $\sigma (x) $ and
$T_{\rm ex}(x)$
are constant but different in the inner and outer regions
following the findings stated in the previous subsection,
and we express them as
\begin{equation}
\label{eq:sigma}
\sigma(x) = \left\{ \begin{array}{l}
{\sigma_{\rm in}}\,\,\,\,(\left| x \right| \le {R_0})\\
{\sigma_{\rm out}}\,\,\,\,({R_0} < \left| x \right|)
\end{array} \right.
\end{equation}
and
\begin{equation}
\label{eq:T}
T_{\rm ex}(x) = \left\{ \begin{array}{l}
{T_{\rm in}}\,\,\,\,(\left| x \right| \le {R_0})\\
{T_{\rm out}}\,\,\,\,({R_0} < \left| x \right|) \,\,.
\end{array} \right.
\end{equation}
For $V_{\rm in}$, we use a sinusoidal function
to mimic the inward motion, i.e.,
\begin{equation}
\label{eq:Vin}
{V_{\rm in}}(x) = \left\{ \begin{array}{l}
V_0 \sin \left( {\frac{{2\pi }}{L}x} \right)\,\,\,\,(\left| x \right| \le L/2)\\
\,\,\,\,\,\,\,\,\,\,\,\,\,\,0\,\,\,\,\,\,\,\,\,\,\,\,\,\,\,\,\,\,\,\,\,(L/2 < \left| x \right|) \,\,
\end{array} \right.
\end{equation}
where $V_0$ and $L$ are constants.

We take $c_{\rm s}$ in Equation (\ref{eq:R0}) to be the sum of thermal 
and turbulent motions, and estimate it
using $T_{\rm ex}(x)$ and $\sigma(x)$ at  $\left| x \right| \le {R_0}$ as
\begin{equation}
\label{eq:cs}
{c_s} = \sqrt {{{\left( {\frac{{k{T_{\rm in}}}}{{{m_{\rm H}}}}} \right)}^2}\left( {\frac{1}{{\overline{\mu}^2}} - \frac{1}{{{\mu ^2}}}} \right) + {\sigma_{\rm in} ^2}}
\end{equation}
where $\mu(=29)$ is the molecular weight of  $^{13}$CO.

We searched for realistic values of the parameters in the above equations,
so that the model can reproduce the observed $^{13}$CO
spectrum in Figure \ref{fig:13CO_C18O}.
By varying $n_0$, $\sigma_{\rm in}$, $\sigma_{\rm out}$, $T_{\rm in}$, $T_{\rm out}$, $L$, and $V_0$
in Equations (\ref{eq:density})--(\ref{eq:cs}) as free parameters,
we fitted the observed $^{13}$CO spectrum with the model
spectrum $T_{\rm e}(V)$ in Equation (\ref{eq:Te}) except for the velocity range
$6.5<V_{\rm LSR}<7.5$ km s$^{-1}$ where the observed spectrum
is contaminated by the bump (``Z"),
and determined the parameters best fitting the data to be
$n_0=3.30^{+1.24}_{-0.84} \times 10^4$ H$_2$cm$^{-3}$,
$\sigma_{\rm in}=0.195^{+0.017}_{-0.018}$ km s$^{-1}$,
$\sigma_{\rm out}=0.457^{+0.062}_{-0.059}$ km s$^{-1}$,
$T_{\rm in}=11.83^{+0.80}_{-0.84}$ K,
$T_{\rm out}=10.58^{+0.51}_{-0.47}$ K,
$L=4.08^{+1.47}_{-0.96}$ pc, and 
$V_0=0.354^{+0.073}_{-0.063}$ km s$^{-1}$.
We show the variations of the parameters along the $x$ axis
in the lower panel of Figure \ref{fig:illust}.
Corresponding values of $R_0$ and $c_s$ in Equations
(\ref{eq:R0}) and (\ref{eq:cs}) are calculated to be $0.51^{+0.12}_{-0.10}$ pc and
$0.26^{+0.02}_{-0.02}$ km s$^{-1}$, respectively.
The total column density estimated as $\int {n(x)dx}$ amounts to
$N$(H$_2$)$= 8.3^{+5.4}_{-3.5} \times 10^{22}$ cm$^{-2}$,
and the velocity gradient at $x=0$ (i.e., $2 \pi V_0 /L$) is $0.55^{+0.29}_{-0.23}$ km s$^{-1}$ pc$^{-1}$.

In Figure \ref{fig:13CO_C18O}(a), we compare
the resulting spectrum of the model (red line) with the
observed $^{13}$CO spectrum (black line).
As seen in the figure, the model can reproduce the
observed spectrum fairly well except for the bump at $V_{\rm LSR}\simeq 7$
km s$^{-1}$.
It is noteworthy that the model can also reproduce the skirts of the observed C$^{18}$O spectrum well
with the model parameters same as for the $^{13}$CO spectrum
but assuming the fractional abundance
$[{\rm C}^{18}{\rm O}]/[{\rm H_2}]=1.7 \times 10^{-7}$ \citep{Frerking1982}
and the molecular weight $\mu=30$,
though the peak intensity of the model spectrum at
$V_{\rm LSR}\simeq 6$ km s$^{-1}$ is much higher than the observed one.
As mentioned earlier, the mismatch is very likely to be caused
by the depletion of C$^{18}$O onto dust in the densest regions of TMC-1.
If we disregard the possible depletion, the derived model parameters
infer the peak (maximum) optical depths of the $^{13}$CO and C$^{18}$O
spectra to be $\sim32$ and $\sim2.7$, respectively. 

Here,
we investigate how each part of the filament should contribute to the observed spectrum.
As shown in Figure \ref{fig:illust}, we divide the inner and outer regions into four regions
$\alpha$, $\beta$, $\gamma$, and $\epsilon$ at $|x|=R_0$,
and calculate the emission from each region in the resulting spectrum.
Results are shown in Figure \ref{fig:13CO_C18O}(b).
The most interesting suggestion inferred from the results is that the emission from the region $\epsilon$
(the outer region closer to the observer) is very similar to the velocity distribution of the additional
component E shown in Figure \ref{fig:13CO_C18O}(a).
This is what we would expect for the component E which should be the diffuse gas lying between
the dense subfilaments A--D and the observer. The results support
the validity of the analyses of the main component of the HC$_3$N line in Section \ref{sec:main}.

Finally, we should note that our model expressed in Equations (\ref{eq:Te})--(\ref{eq:cs}) is
a simplified toy model made only for the purpose to see whether the observed $^{13}$CO and C$^{18}$O spectra
in Figure \ref{fig:13CO_C18O} can be reproduced by the global inward motion of the filament.
In the above analyses, we assume the density distribution of an isothermal and gravitationally stable filament.
However, velocity dispersion and temperature in TMC-1 should be more complex than we assumed and vary in reality. 
Also, there is no proof that the filament is gravitationally stable, and
the density profile $n(x)$ should not follow Equation (\ref{eq:density}) precisely.
Actually, as we will show in Section \ref{sec:exist}, the density profile in the dense inner region
traced in CCS may be different from what is inferred from the $^{13}$CO data in this subsection.
In addition, we should note that other density distributions such as those for gravitationally stable ($n \propto |x|^{-2}$) or
dynamically infalling ($n \propto |x|^{-1.5}$) spherical clouds can also reproduce similar spectra.
The point of the analyses in this subsection is that, in order to reproduce the observed spectra,
we need to assume the global inward motion of the filament which has a centrally peaked density profile
and slightly warmer and less turbulent interior than the outer region.
Under these assumptions, the unseen component E inferred from the analyses
of the HC$_3$N spectrum (Section \ref{sec:main})
can also be naturally understood as the diffuse molecular gas
located between TMC-1 and the observer (i.e., the part $\epsilon$ in Figure \ref{fig:illust}).

\subsection{Gravitational Stability}

All of the above nice matches of the observations and the model lead us to the conclusion
that the entire TMC-1 system is moving inward.
Here, we should note that the inward motion inferred from the $^{13}$CO and C$^{18}$O spectra
does not necessarily mean the ``dynamical infalling motion" or ``collapsing motion" of the filament
by the self-gravity leading to star formation.
In order to
discuss
whether TMC-1 is globally collapsing by the self-gravity or not, we will estimate the Virial mass and
compare it with the mass of the filament.

If we disregard the contributions of the external pressure and the magnetic field,
the mass of the filament per unit length that can be supported by the
gas pressure
is estimated as
\begin{equation}
\label{eq:Mvir}
{M_{\rm vir}} = \frac{{2{c_{\rm s}^2}}}{G} = 464{\left( {\frac{{{c_{\rm s}}}}{{\rm km\,{s^{ - 1}}}}} \right)^2} \,\,M_{\sun} \, {\rm pc^{ - 1}} \, .
\end{equation}
If we take $c_{\rm s}=0.26$ km s$^{-1}$ as derived in the previous subsection,
we get $M_{\rm vir}\simeq 30$ $M_{\sun}$ pc$^{-1}$.
The Virial mass should decrease to half of this value at most ($M_{\rm vir}\simeq15$ $M_{\sun}$ pc$^{-1}$)
if the filament is surrounded by a high external pressure, which is often seen among small dark
clouds with a mass of a few hundred solar masses
\citep[e.g.,][]{Dobashi1996,Dobashi2001,Yonekura1997,Shimoikura2013,Shimoikura2018}.

On the other hand, the C$^{18}$O map in Figure \ref{fig:13co_c18o_maps} infers
the molecular mass and the length of TMC-1 of $\sim200$ $M_\sun$
and $\lesssim1$ pc, indicating that the actual mass of the filament per unit length $M_{\rm c}$ is
an order of $\gtrsim 200$ $M_\sun$ pc$^{-1}$, which is several times larger than $M_{\rm vir}$.
Also, mass of the filament estimated from the fitted values of
$R_0$($\simeq0.51^{+0.12}_{-0.10}$ pc) and $n_0$($\simeq3.30^{+1.24}_{-0.84} \times 10^4$ H$_2$cm$^{-3}$) as
${M_{\rm c}} = \mu {m_{\rm H}}{n_0}\pi R_0^2$
\citep{Stodolkiewicz1963,Ostriker1964,Inutsuka1992}
is 
$1.9^{+2.1}_{-1.0} \times 10^3$ $M_\sun$ pc$^{-1}$
which is much higher than $M_{\rm vir}$, 
though this mass estimate should suffer from a large uncertainty,
because the real density $n(x)$ may not follow Equation (\ref{eq:density})
precisely as mentioned at the end of the previous subsection.

In any case, the condition $M_{\rm c} > M_{\rm vir}$ is very likely to hold,
which means that TMC-1 should be contracting by its self-gravity,
if there is no supporting force other than the gas pressure.
In that case, the inward motion inferred from our analyses should
represent the infalling motion by the self-gravity of the filament. 

However, we wonder if TMC-1 is really contracting, because it would soon form
a star as the derived $n_0$ infers a short free-fall time of
$\tau_{ff} \simeq 2 \times10^5$ year, whereas there have not been found young
stellar objects around the observed position.
A very likely possibility is the supporting force by the magnetic field.
A cloud with a mass and size of TMC-1 ( $10^2$ $M_\sun$ and $\sim1$ pc)
could be easily supported by the magnetic field of an order of $B \sim 10^2$ $\mu$G
\citep[e.g.,][]{Nakano1985,Shu1987}, and we actually found that the magnetic field
in TMC-1 is very likely to be around $\sim10^2$ $\mu$G through the analyses
of the Zeeman splitting of the CCS line (Nakamura et al. 2018, in preparation).

Though the Virial analyses infer that TMC-1 can contract by the self-gravity,
we would rather suggest that TMC-1
is globally in the gravitational equilibrium if we take into account 
the contribution of the magnetic field,
and that it is oscillating around the equilibrium point,
like what has been proposed to the isolated globule B68 \citep{Lada2003}.

\subsection{Do the Subfilaments Exist?}\label{sec:exist}

\begin{figure}
\begin{center}
\includegraphics[scale=0.8]{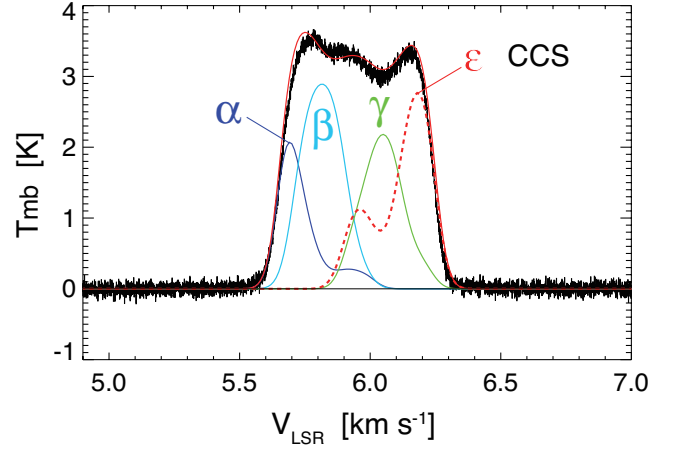}
\caption{
The observed CCS spectrum (black line)
compared with the simulated spectrum (red solid line) based on the model in Figure \ref{fig:illust}.
Model parameters are set to
$L=0.5$ pc,
$V_0=0.245$ km s$^{-1}$,
$n_0=1\times10^5$ cm$^{-3}$,
$\sigma_{\rm in}=0.05$ km s$^{-1}$,
$\sigma_{\rm out}=0.05$ km s$^{-1}$,
$T_{\rm in}=7.4$ K, and
$T_{\rm out}=6.0$ K.
We assumed the fractional abundance of CCS to be $[{\rm CCS}]/[{\rm H_2}]=5 \times 10^{-8}$.
The set of the parameters yields the peak (maximum) optical depth and radius
of $\tau \simeq1.9$ and $R_0=0.17$ pc.
Contributions of the four parts labelled $\alpha$, $\beta$, $\gamma$, and $\epsilon$
in Figure \ref{fig:illust} but divided at $|x|=0.5R_0$ instead of  $|x|=R_0$
where $R_0 \simeq 0.17$ pc are shown by the lines with different colors.
\label{fig:model_ccs}}
\end{center}
\end{figure}

On the way of the analyses of the $^{13}$CO spectrum in Figure \ref{fig:13CO_C18O}(b), 
we realized that the model in Figure \ref{fig:illust} can produce a spectrum similar to the CCS line,
if we tune the model parameters.
An example is shown in Figure \ref{fig:model_ccs}
where we set the parameters in Equations (\ref{eq:density})--(\ref{eq:cs}) to
$n_0=1 \times 10^5$ H$_2$cm$^{-3}$,
$\sigma_{\rm in}=\sigma_{\rm out}=0.05$ km s$^{-1}$,
$T_{\rm in}=7.4$ K,
$T_{\rm out}=6.0$ K,
$L=0.5$ pc, and
$V_0=0.245$ km s$^{-1}$.
In addition, we set the molecular weight in Equation (\ref{eq:cs}) and 
the fractional abundance of CCS to be $\mu=56$ and
$[{\rm CCS}]/[{\rm H_2}]=5 \times 10^{-8}$.
The value of $[{\rm CCS}]/[{\rm H_2}]$ giving a good fit to the observed spectrum
is unexpectedly large being higher than those measured in any other molecular clouds
by a factor of $\gtrsim10$ \citep[e.g.,][]{Shimoikura2018}, though such
a high fractional abundance of CCS could be possible for
molecular clouds in a very early stage of evolution \citep[e.g.,][]{Suzuki1992}.
In this subsection, we shall temporarily use the value without examining its legitimacy,
because we cannot find a more plausible set of parameters at once.
The set of the above parameters corresponds to the effective radius $R_0\simeq0.17$ pc
and the maximum optical depth of the CCS line $\tau \simeq 1.9$.

As seen in Figure \ref{fig:model_ccs}, the overall shape of the model CCS spectrum is very similar
to the observed one. Although we don't attempt to optimize the model parameters
because we are not confident of the assumed $[{\rm CCS}]/[{\rm H_2}]$ value,
the adopted parameters such as $\sigma_{\rm in}$, $\sigma_{\rm out}$, $T_{\rm in}$, and $T_{\rm out}$
are likely to be significantly different from those of the $^{13}$CO and C$^{18}$O lines derived
in the previous subsection,
but are closer to those of the velocity components A--D measured in CCS (Table \ref{tab:ccs}).
This may infer that the global distributions of the parameters are complex and cannot be
expressed precisely by the simple relations in Equations (\ref{eq:density})--(\ref{eq:Vin})
over the wide range of density from diffuse outskirts (tranced in CO isotopologues)
to the densest central region (traced in CCS).

In Figure \ref{fig:model_ccs}, contributions of the four regions $\alpha$ -- $\epsilon$ in Figure \ref{fig:illust}
(but divided at $|x|=0.5R_0$ instead of $|x|=R_0$)
are also shown by the different colors.
It is very interesting that the emission lines from the four regions also appear similar
to the velocity components A--D that we found in Section \ref{sec:results}
(see Figure \ref{fig:ccs}).
The similarity infers an important possibility that the velocity components A--D which we regard
as the subfilaments in this paper might merely represent the emission from the four regions
$\alpha$ -- $\epsilon$ of a single filament which doesn't have any subfilaments.

In this paper, we regard that the velocity components A--D
represent the real subfilaments or substructures in TMC-1, not merely
representing different parts of a single continuous filament.
This is mainly because of the following two reasons:
(1) The optically thin satellites of the HC$_3$N
line are apparently asymmetric with respect to the systemic
velocity (Figure \ref{fig:F4455}), while the total emission from
the regions $\alpha$ -- $\epsilon$ in Figure \ref{fig:illust} should
appear symmetric in the optically thin lines
for any sets of the model parameters, and
(2) further analyses of the CCS data on a large scale (Figure \ref{fig:intensity_maps})
infer that the velocity components A--D are likely to coincide
with four elongated subfilaments (Dobashi et al. 2018, in preparation).

However, concerning the above point (1), we should note that our model
in Figure \ref{fig:illust} assumes
ideal symmetric distributions of the model parameters
(density, temperature, and inward velocity) along the $x$ axis,
but their true distributions may not satisfy our assumptions.
It could be also possible that the satellite lines of HC$_3$N can be
optically thick but distribute in a patchy way within the beam of the 45m telescope.
In that case, the satellite lines would show asymmetric profiles,
though they would appear faint and optically thin due to the beam dilution.
Also, concerning the point (2), the four elongated subfilaments
are somewhat aligned and located near the center of TMC-1,
inferring that they might represent the oscillation of the main filament.

To summarize, we regard in this paper that the four velocity components A--D originate from
physically distinct subfilaments, but we cannot entirely rule out the possibility that
they might be representing different parts of a single continuous filament without any substructures.
In order to determine which case is realistic, it is crucial to resolve the spatial
distributions of the satellites of the HC$_3$N line at a very high angular resolution that can be
achieved by a large interferometer such as ALMA.


\section{CONCLUSIONS}\label{sec:conclusions}

We have observed the cyanopolyyne peak in the TMC-1 filament
with the CCS($J_N=4_3-3_2$) and HC$_3$N($J=5-4$) emission lines at 45 GHz utilizing the
Z45 receiver and PolariS spectrometer installed in the 45m telescope
at the Nobeyama Radio Observatory (NRO). Thirty-hours integration with
these instruments provided us with wonderful spectral data having
a very high velocity resolution of 0.0004 km s$^{-1}$ and a noise
level of 0.04 K. Analyses of the spectra infer various information
on the structure of TMC-1 along the line-of-sight (LOS). Main conclusions
of this paper are summarized in the following:

\begin{enumerate}

\item
Based on the analyses of the optically thin $F=4-4$ and $5-5$ hyperfine lines of the HC$_3$N emission,
we identified four velocity components along the LOS of TMC-1.
These components are separated by $0.1-0.2$ km s$^{-1}$ and have a line width of $0.05-0.09$ km s$^{-1}$.
We regard that the four velocity components represent subfilaments in the TMC-1 filament, and
we named them A, B, C, and D in the order of increasing radial velocity.

\item
To investigate the order of the velocity components along the LOS,
we solved radiative transfer for the CCS and HC$_3$N spectra
taking into account the effect of absorption by the components lying
in the foreground of the other components. Results infer that the four components
are located in the order of A, B, C, and D from the further to closer positions
to the observer. The fact that the components with lower radial velocity
are lying at further positions from the observer indicates that these velocity
components are moving inward toward the center of the TMC-1 filament.

\item
Results of the analyses of the optically thick $F=6-5$, $5-4$, and $4-3$ hyperfine lines of the HC$_3$N
emission infer the existence of an additional component which should be diffuse gas
lying in the foreground of the other components and contribute to the observed HC$_3$N spectrum
only as the absorber without emitting the molecular line by itself.
The component that we call E follows the same velocity trend as the other components A--D.

\item
We investigated the $^{13}$CO and C$^{18}$O spectra downloaded from the data archive of NRO,
and found an evidence of the global inward motion of the TMC-1 filament.
To better understand the global structure of the filament, we made a simple model
which can account for the main features of the $^{13}$CO and C$^{18}$O spectra
as well as the origin of the additional component E.

\item
Virial analyses infer that the TMC-1 filament can collapse by the self-gravity
within a time scale of $\sim2 \times 10^5$ years,
unless it is supported by the magnetic field of an order of $\sim100$ $\mu$G.
Because there is no YSO forming there, we suggest that TMC-1 is 
in the gravitational equilibrium being supported by the magnetic field,
and that the observed inward motion may
represent oscillation of the filament.

\item
Our model infers a possibility that the observed velocity components A--D might be
the emission from different parts of a shrinking
or oscillating
single filament
which does not necessarily need subfilaments.
In the case of TMC-1, we regard that the four components represent real
subfilaments,
because they show
asymmetric velocity distributions in the optically thin lines
with respect to the systemic velocity.
However, in oder to check how much this picture is realistic,
it is crucial to resolve TMC-1 at a very high angular resolution that can be
achieved by a large interferometer such as ALMA.

\end{enumerate}


\acknowledgments
We thank Tetsu Ochiai, Jun'ichi Hirahara, and Atsumi Goto for their support
of the observations and analyses.
This work was financially supported by Grant-in-Aid for Scientific Research 
(Nos. 17H02863, 17H01118, 26287030, 17K00963)
of Japan Society for the Promotion of Science (JSPS). 
\clearpage




\begin{deluxetable*}{cclcccrc} 
\tablecaption{The Observed Molecular lines \label{tab:45obs}} 
\tablehead{ 
 \colhead{Molecule} & \colhead{Transition}   & \colhead{Rest Frequency}  & \colhead{Line Intensity}  & \colhead{$j$} & \colhead{$R_j$}  & \colhead{$V_f^j$} &  \colhead{Comment}\\
 \colhead{} &\colhead{}  & \colhead{(MHz)} & \colhead{} & \colhead{} &  \colhead{ }  &  \colhead{(km s$^{-1}$)}  &  \colhead{ }  
}
\startdata 
CCS			&	$J_{N}$=4$_{3}-3_{2}$	&	45379.04600 (0.02)		&	3.972	& ... &	...	&	...	&	...	\\
HC$_{3}$N	&	$J=5-4$, $F=5-5$		&	45488.83680 (0.001)		&	0.067	& 0 &	0.0134	&	9.7331	&	satellite\\
HC$_{3}$N	&	$J=5-4$, $F=4-3$		&	45490.26140 (0.0005)	&	1.296	& 1 &	0.2592	&	0.3447	&	main\\
HC$_{3}$N	&	$J=5-4$, $F=5-4$		&	45490.31370 (0.0005)	&	1.600	& 2 &	0.3200	&	0.0000	&	main	\\
HC$_{3}$N	&	$J=5-4$, $F=6-5$		&	45490.33730 (0.0006)	&	1.970	& 3 &	0.3940	&	-0.1555	&	main	\\
HC$_{3}$N	&	$J=5-4$, $F=4-4$		&	45492.10850 (0.0009)	&	0.067	& 4 &	0.0134	&	-11.8282	&	satellite\\
\enddata 
\tablecomments{The rest frequencies and line intensities are taken from Splatalogue. Numbers in the parentheses for the rest frequencies
are the cataloged ambiguities. For the HC$_3$N line, we call the three blended and two isolated hyperfine
lines ``main'' (for $j=1, 2, 3$ in the table) and ``satellite'' (for $j=0$ and $4$) as denoted in the column for comments.
$R_j$ is the normalized line intensities of the HC$_3$N lines.
$V_f^j$ is the velocity shift relative to the $F=5-4$ line of HC$_3$N (see text).}

\end{deluxetable*} 



\begin{deluxetable*}{cccc} 
\tablecaption{Model Parameters for the HC$_3$N Satellite Line \label{tab:satellite}} 
\tablehead{ 
	\colhead{Component} & \colhead{$T_{\rm s}^i$}  &	\colhead{$V_0^i$}	&	\colhead{${\sigma^{i}}$} \\
	\colhead{ } 	&  \colhead{(K)} 	&	\colhead{(km s$^{-1}$)} 	&	\colhead{(km s$^{-1}$)} 	
}
\startdata
A	&	0.480(0.024) 	&	5.7271(0.0018) 	&	0.0542(0.0010) 	\\
B	&	0.598(0.015) 	&	5.9014(0.0065) 	&	0.0881(0.0091) 	\\
C	&	0.346(0.078) 	&	6.0636(0.0065) 	&	0.0613(0.0116)	 	\\
D	&	0.308(0.072) 	&	6.1602(0.0061) 	&	0.0474(0.0023) 	\\
\enddata
\tablecomments{Model parameters in Equation (\ref{eq:satellite}) best fitting the observed HC$_3$N satellite spectra.
Numbers in the parentheses are the standard error of the parameters.}

\end{deluxetable*}


\begin{deluxetable*}{cccccc} 
\tabletypesize{\scriptsize}
\tablecaption{Model Parameters for the CCS Line \label{tab:ccs}} 
\tablehead{ 
	\colhead{Component} & \colhead{$T_{\rm 0}^i$}  &\colhead{$V_0^i$} & \colhead{${\sigma^i}$} & \colhead{$\tau_{\rm 0}^i$}  & \colhead{$T_{\rm ex}^i$} \\
	\colhead{ } 	&  \colhead{(K)} 	&	\colhead{(km s$^{-1}$)} 	&	\colhead{(km s$^{-1}$)}   &   \colhead{ } 	&  \colhead{(K)}
}
\startdata 
A	&	2.223(0.044) 	&	5.7271 	&	0.0523(0.0006) 	&	2.152(0.072)	&	5.98(0.11) \\
B	&	3.169(0.074) 	&	5.9014 	&	0.1155(0.0013) 		&	1.367(0.066)	&	7.34(0.17)\\
C	&	2.195(0.013) 	&	6.0636	&	0.0703(0.0020) 	&	1.876(0.101)	&	5.94(0.09)\\
D	&	2.785(0.031) 	&	6.1602	&	0.0465(0.0003) 	&	1.084(0.047)	&	6.79(0.12) \\
\enddata

\tablecomments{Model parameters in Equations (\ref{eq:radiative_trans}) and (\ref{eq:tau_ccs}) best fitting the observed CCS spectrum.
Numbers in the parentheses stand for the standard error of the parameters.
The best value and the error of the other parameter $V_{\rm shift}$ in Section \ref{sec:ccs}
is $V_{\rm shift}=$0.0158(0.0010) km s$^{-1}$. For $T_{\rm ex}$, see Section \ref{sec:Tex}.}
\end{deluxetable*} 


\begin{deluxetable*}{cccccc} 
\tablecaption{Model Parameters for the HC$_3$N Main Line \label{tab:main}} 
\tablehead{ 
	\colhead{Component} & \colhead{$T_0^i$}  &\colhead{$V_0^i$} & \colhead{${\sigma^i}$} & \colhead{${\tau_0^i}$} & \colhead{$T_{\rm ex}^i$} \\
	\colhead{ } 	&  \colhead{(K)} 	&	\colhead{(km s$^{-1}$)} 	&	\colhead{(km s$^{-1}$)}   &   \colhead{ }  &   \colhead{(K)} 	
}
\startdata
A	&	3.114(0.020) 	&	5.7271 	&	0.0542 	&	12.50(0.60)  	&	7.26(0.13)\\
B	&	5.103(0.031) 	&	5.9014 	&	0.0881 	&	9.30(0.23)  	&	10.12(0.21)\\
C	&	6.254(0.043) 	&	6.0636 	&	0.0613 	&	4.25(0.93)  	&	11.77(0.26)\\
D	&	5.001(0.029) 	&	6.1602 	&	0.0474 	&	4.75(1.08) 	&	9.97(0.21) \\
E	&	... 			&	6.2147(0.0045) 	&	0.5497(0.0031) 	&	0.582(0.006) &	...\\
\enddata
\tablecomments{Model parameters in Equations (\ref{eq:radiative_trans}) and (\ref{eq:tau_hc3n}) best fitting
the observed HC$_3$N main spectrum.
Numbers in the parentheses are the standard error of the parameters. Except for the component E,
we left only $T_0^i$ as the free parameters, and fixed $V_0^i$ and $\sigma^{i}$ to the values in Table \ref{tab:satellite}.
We also calculated $\tau_0^i$ and the error from the fitted $T_0^i$ and $T_{\rm s}^i$ following Equation (\ref{eq:estimate_tau_m0}).
For $T_{\rm ex}$, see Section \ref{sec:Tex}.
}
\end{deluxetable*}


\clearpage




\end{document}